\documentclass{article}

\usepackage[preprint, nonatbib]{nips_2018}

\usepackage[skip=2pt]{caption}
\title{A Scalable, Flexible Augmentation of the Student Education Process}

\author{Bhairav Mehta \\
  Massachusetts Institute of Technology\\
  {\tt bhairavm@mit.edu}
  \And
  Adithya Ramanathan \\
  University of Michigan \\
  {\tt adithram@umich.edu}}
\date{}

\begin{document}

\maketitle

\begin{abstract}
  We present a novel intelligent tutoring system which builds upon well-established hypotheses in educational psychology and incorporates them inside of a scalable software architecture. Specifically, we build upon the known benefits of knowledge vocalization \cite{Ausubel1961-AUSIDO-2}, parallel learning \cite{Topping1996}, and immediate feedback \cite{samuels_wu} in the context of student learning. We show that open-source data combined with state-of-the-art techniques in deep learning and natural language processing can apply the benefits of these three factors at scale, while still operating at the granularity of individual student needs and recommendations. Additionally, we allow teachers to retain full control of the outputs of the algorithms, and provide student statistics to help better guide classroom discussions towards topics that would benefit from more in-person review and coverage. Our experiments and pilot programs show promising results, and cement our hypothesis that the system is flexible enough to serve a wide variety of purposes in both classroom and classroom-free settings.
\end{abstract}

\section{Introduction}

One notable constant throughout recent history has been the process of education. Despite changing mediums from paper to pixels, the \textit{structure} of education has stayed nearly identical: sequential, isolated learning, bookended by exams and benchmarked by grades. Yet incredibly, even this system has not propagated through all levels of society - many neighborhoods, regions, or entire countries still lack access to quality education \cite{UNSDG}. While many studies have been done about the factors that are conducive to learning, there are few solutions that scale well. Educators are often aware of the issues, but constrained by combinations of time, resources, or attention, they are forced to adopt a one-size-fits-all system for education. 

In this paper, we develop a novel framework that harnesses open-source data and state-of-the-art deep learning and natural language processing techniques to replicate key factors for effective learning, as cited by educational psychologists. Specifically, when students vocalize what they have learned, receive feedback, and learn about related topics when struggling with the main one, both knowledge recall (short-term) and retention (long-term) has been proven to increase \cite{Topping1996}. Our main contribution can be summarized as applying these beneficial factors at scale with software, without significant increases in time investment from educators.

We develop an end-to-end framework that asks students to vocally answer questions created by their teacher, who provided pertinent data sources when creating the question. We then perform semantic similarity tests between a student answer and the key concepts extracted from source data, and provide feedback to the student about their performance. When a student is struggling with a concept, we recommend other questions - either similar ones also created by the teacher or ones automatically generated from the data sources relevant to the topic -  helping plug gaps in knowledge.

On the educator side, our system allows teachers to quickly create questions, needing only to provide sources of information before the algorithms extract key concepts from the data. We give teachers the final say in any decisions made by the system, ensuring that every machine output is screened and approved by a human expert. We stay true to the unofficial education-technology motto of minimizing teacher investment while maximizing student impact \cite{norris_soloway02/12/18}, and develop a pipeline that uses machine learning to augment the education process. We show empirical evidence that providing teachers with summarized statistics about student performance can help increase the efficiency of the class time by focusing discussion on difficult or confusing topics.

Finally, we extend our framework even further to show that such a system can also be used independently from a classroom, and lay the groundwork to enable efficient learning in places that may not have access to formal education. 

\subsection{Related Work}
Education technology (ET) is defined as the "practice of facilitating learning [by] creating, using, and managing appropriate technological processes" \cite{hlynka_jacobsen}. Many ET solutions do not make good use of either the educational infrastructure already in place, or the expertise of teachers in the classrooms. Leading ET applications offer deployment flexibility to the educator, constant interaction with the student, and ease of setup through pre-built curricula. Yet, many of these successful apps are still constrained by the fact that their materials are entirely expert curated. Our approach builds upon the principles of interaction and feedback, but goes one step further. We offer educators a way to intelligently and quickly build custom curricula and provide recommendations to struggling students by using only a few teacher-provided data sources.


\section{Implementation}

Our system can be broken down into three main sections, depending on which user is interacting with the system. \textit{Pre-Teacher Setup} describes data collection done to provide the groundwork for the concept extraction algorithms, and \textit{Teacher Usage} describes the process a teacher goes through to generate questions for students to answer. Finally, \textit{Student Usage} details the mechanism for both answering questions and receiving recommendations on other topics or questions to study.

\subsection{Pre-Teacher Setup} \label{pre_teacher}
For each subject area supported by our framework, we create a tf-idf index \cite{Ramos_usingtf-idf} using a hand-crafted list of seed URLs that we believe are pertinent to the topic (i.e the subject of US History may have Wikilinks to different eras in American History). We use a web-spider to crawl the seed links to a configurable depth, and extract the text from each unique link as a separate document. We use standard text preprocessing techniques \cite{DBLP:journals/corr/AllahyariPASTGK17a} in the creation of these indices, and mitigate the issue of domain transfer by maintaining multiple subject-based tf-idf indices. In parallel, we train a standard Paragraph Vector \cite{DBLP:journals/corr/LeM14} model with the GenSim Python Library \cite{rehurek_lrec} using the cleaned data, and store only the encoder for use in the student recommendation process described in Sec. \ref{student}.

\subsection{Teacher Usage} \label{teacher}
Upon creating a class, the teacher selects a subject area that roughly corresponds to the class being taught, which, in the background, links a relevant tf-idf index to the class. The teacher, now creating questions, is asked to provide two pieces of information per question: a Question Title (seen by the student when asked to answer), and Data Sources. The data sources, links or blocks of text, are assumed to be relevant to the question and material covered in class, and are then preprocessed using the same standard techniques as in Sec. \ref{pre_teacher}. Once we have cleaned text, we use a LSTM-CRF model \cite{DBLP:journals/corr/LampleBSKD16}, trained on the Annotated Groningen Meaning Base \cite{Bos2017GMB}, to extract named entities (NE). In parallel, we run the text through a TextRank model \cite{Mihalcea04TextRank, DBLP:journals/corr/BarriosLAW16} to retrieve a list of key concepts (KC) from the source. We calculate the score for each word in the raw text using a weighted score of its tf-idf weight, and indicator functions corresponding to its presence in either the NE or KC lists.

\begin{equation}\label{scoreeqn}
    s(w) = TF(w) + \alpha I[KC(w)] + \beta I[NE(w)]
\end{equation}

where $TF(w)$ is the tf-idf score of the word, $I$ is the indicator function depicting whether a word is a member of the set, and $\alpha$ and $\beta$ are empirically-set hyperparameters weighing the contribution of the Key Concept and Named Entity scores. If appropriate, we combine the words into phrases using the two lists of NEs and KCs, as well as the NLTK Multiword Expression Tokenizer \cite{Loper02nltk:the}, and assign the phrase the sum of its component scores. The phrases and words, called \textit{concepts}, are returned to the teacher in a truncated list of \texttt{(concept, score)} pairs, where both concepts and score values can be manually adjusted by the educator. We also embed the raw text extracted from the question's associated data using the trained Paragraph Vector model described in Sec. \ref{pre_teacher}, and store the "question embedding" in a database.

While most of our recommendations for struggling students come from similarity tests between the stored question embeddings, we also experiment with neural question generation, whose outputs we provide to the teacher alongside the \texttt{(concept, score)} pairs. To do this, we train a Paragraph Level question-generation model as described in \cite{du2017learning}, and feed it the five sentences that had the highest total sentence score, where score was computed by summing the score given by all of the \texttt{(concept, score)} pairs in a given sentence. While many of these questions were not approved by the teacher to be shown to the student, we highlight this as an exciting area to try out newer, more sophisticated question-generation models such as those described in \cite{DBLP:journals/corr/YuanWGSBSZT17}.

\subsection{Student Usage}\label{student}
When a student enrolls in a class, he will see the questions proposed by the teacher, only with the question title. The student enables his microphone, and answers the question by speaking to the computer using classroom or background knowledge. To handle speech-to-text, we utilize Mozilla's open source implementation of DeepSpeech \cite{DBLP:journals/corr/HannunCCCDEPSSCN14, MozillaDeepSpeech}. We preprocess the transcripted text with the same methods used in Sec. \ref{pre_teacher}. From there, we tokenize the answer, and then score it using Equation \ref{scoreeqn}. We contribute the phrase's full score even for partial hits with student answers, and acknowledge that a more sophisticated scoring scheme could be used. 

Along with the score and a visual representation of which words in his answer matched up with key concepts associated with the question, we also provide the student recommendations to other similar questions to enable parallel learning. We compute the cosine similarity between the embeddings of the current question and all other questions created by the teacher within the particular class, and return the questions that are the three nearest neighbors in embedding space. As the embedding space ideally captures semantics, the recommendations are questions whose data sources relate to the current topic's data sources, allowing the student to tackle the same subject area from various different angles. Semantic embedding of data sources is helpful here, because even if a student may not believe the question titles are related, the answers themselves might be. This allows for students to find connections between seemingly disparate topics and learn faster on both fronts. 

In addition, from our initial experiments, we find that many users utilize pronouns extensively. Given our scoring formulation, the use of pronouns at times leads to misses between the user response and the target response. To help mitigate this issue, we utilize co-reference resolution techniques to resolve these referential pronouns and mentions within user responses. Coreference resolution detects mentions within a body of text, and then analyzes the statement for possible antecedents for the given mention. Resolving pronouns allows us to more accurately score longer answers, where we empirically see the increased usage of pronouns. We deploy models similar to those described in \cite{clarkmanning-acl16} and \cite{clarkmanning-emnlp16}, and replace each pronoun with its predicted antecedent in the response before scoring.


\section{Results}
For individual elements of the pipeline, such as the LSTM-CRF or the Paragraph Vector networks, we use many of the same training datasets and cross-validation procedures as described in the original papers.  
We also conducted a user study to evaluate system performance (relevance of key concepts extracted and relevance of recommended questions) on 10 example questions, judged by 24 respondents (both teachers and students). We asked for relevance ratings (1-5 scale, 5 being of high relevance) for each of the key concepts and recommended questions, and present truncated results in Table \ref{user-table}. We also gauge the importance and effectiveness of the three main embodiments of the effective learning hypotheses within our system: \textbf{(A)} Knowledge Vocalization, \textbf{(B)} Parallel Learning (via Recommendations), \textbf{(C)} and Immediate and Visual feedback. Lastly, we asked teachers for ratings describing the \textbf{(D)} Student Performance Statistics and present truncated results using a similar scale in Table \ref{embodiment-table}.
\begin{table}[h!] \label{study}
\def\arraystretch{1.15}
\centering
\begin{tabular}{|c|c|c|c|c|c|}
    \hline \bf Question & \bf Q1 & \bf Q7 & \bf Q10\\  
    \hline
    \bf Avg. Relevance of KC  & 4.33 & 3.50 & 4.00 \\ 
    \bf Avg. Relevance of Rec. ?s &  1.33 & 3.20 & 4.50 \\ 
    \hline 
\end{tabular}
\caption{\label{user-table} Sampled relevance scores, averaged across 24 respondents}
\begin{tabular}{|c|c|c|c|c|c|}
    \hline & \bf (A) & \bf (B) & \bf (C) & \bf (D) \\  
    \hline
    \bf Component Importance & 3.75 & 4.33 & 2.50 & 5.00 \\ 
    \bf Component Effectiveness &  4.25 & 2.33 & 2.33 & 4.33 \\ 
    \hline
\end{tabular}
\caption{\label{embodiment-table} Importance and Effectiveness Scores, averaged across 24 respondents}
\end{table}
Our most surprising conclusions came from pilot programs conducted in an American school system. The program used web-based implementation of the system inside of three classrooms studying various Advanced Placement (AP) subjects. All three teachers, given just the system, were using it in entirely different ways: one as a complete homework substitute, one as a homework add-on, and another as a independent (i.e no teacher enforcement of usage) self-study tool for annual AP exams. These results show that even a barebones implementation of this framework, one driven by open-source information sources and widely available machine learning algorithms, can be easily molded to varying use cases with almost no extra effort from educators. 

In addition, Table 2 summarizes our discussions with the pilot program's teachers, which showed that the system's class performance statistics (labeled (\textbf{D})) were overwhelmingly described as the most important feature. Class-level feedback allowed for more focused discussions during class time with students. During the short pilot test, we were told that teachers' day-to-day schedules became more dynamic, and time devoted to discussing each topic was conditioned on the performance of the students, rather than allocated from the teacher's intuition of where students face difficulty.

\section{Classroom-Free Extension}

We now describe two avenues to extend our system in a classroom-free setting. First, content creators can create curricula that others can use, improve, and share, which is a similar model driving many popular Massive Open Online Courses. Our pilot program respondents describe that, with a small amount of prior research and data gathering, an entire class curriculum can be created in under a day. 
More excitingly, the curriculum generation can also be automated, sharing much of the infrastructure that drives the classroom-based system. In this route, we ask the student for the relevant data sources (text and links), and then perform the same concept extraction and answer scoring as described in Sec. \ref{teacher} and Sec. \ref{student}. Importantly, when providing sources to the system, the student only needs to know \textit{what} is relevant, and the system will ideally help him learn \textit{why}. We then create and pose automatically-generated questions to the student as described in Sec. \ref{teacher}.

While this method showed poor results in early tests, presumably for the same reason that many generated questions were not approved by teachers, we believe that more sophisticated question-generation methods will be able to extend the benefits of this method to help people learn \textit{any} new concept.

\section{Conclusion}
We present a simple yet effective system that builds on popular hypotheses in educational psychology, and reproduce them in a scalable software framework. We provide a flexible solution that can serve a wide variety of purposes in a classroom, while keeping educators in the loop every step of the way. We present empirical evidence that backs the claims we make about effective learning, and show preliminary results for a way to extend the system to places or people that have do not have access to a formal education. As the fields of education and machine learning move forward (hopefully together), our results show that the most important area of focus may be in automatic question generation. The right mix of machine learning models can provide students and teachers enormous impact when turned towards an education setting, and we hypothesize that many of the benefits would remain in a classroom-free implementation as well.




\bibliography{nips_2018} 
\bibliographystyle{abbrv}

\end{document}